# Field emission from AlGaN nanowires with low turn-on field

Filippo Giubileo *,[a], Antonio Di Bartolomeo *,[b,a], Yun Zhong [c], Songrui Zhao [c], Maurizio Passacantando [d]

[a] CNR-SPIN Salerno, via Giovanni Paolo II n. 132, Fisciano 84084, Italy.
[b] Physics Department, University of Salerno, via Giovanni Paolo II n. 132, Fisciano 84084, Italy.
[c] Department of Electrical and Computer Engineering, McGill University, 3480 University Street, Montreal, Quebec H3A 0E9, Canada
[d] Department of Physical and Chemical Science, University of L'Aquila, and CNR-SPIN L'Aquila, via Vetoio, Coppito 67100, L'Aquila, Italy.

E-mail: filippo.giubileo@spin.cnr.it (Filippo Giubileo) and adibartolomeo@unisa.it (Antonio Di Bartolomeo).

**Abstract**

We fabricate AlGaN nanowires by molecular beam epitaxy and we investigate their field emission properties by means of an experimental setup using nano-manipulated tungsten tips as electrodes, inside a scanning electron microscope. The tip-shaped anode gives access to local properties, and allows collecting electrons emitted from areas as small as 1µm$^2$. The field emission characteristics are analysed in the framework of Fowler-Nordheim theory and we find a field enhancement factor as high as β = 556 and a minimum turn-on field $E_{turn-on}$ = 17 V/µm for a cathode-anode separation distance $d$ = 500 nm. We show that for increasing separation distance, $E_{turn-on}$ increases up to about 35 V/µm and β decreases to ~100 at $d$ = 1600 nm. We also demonstrate the time stability of the field emission current from AlGaN nanowires for several minutes. Finally, we explain the observation of modified slope of the Fowler-Nordheim plots at low fields in terms of non-homogeneous field enhancement factors due to the presence of protruding emitters.

## 1. Introduction

Binary group III-Nitride semiconductors, such as GaN, AlN, and InN are important materials for optoelectronic applications, due to their direct bandgap and doping capabilities [1]. The large breakdown field, the high electron mobility, and their mechanical and thermal robustness, make them optimal candidates also for high power electronics [2]. In the last two decades, one-dimensional (1D) III-V semiconductor nanowires (NWs) have attracted great attention due to several unique physical, electrical and optical properties [3] that can be exploited for next-generation devices such as solar cells [4], light emitting diodes [5], memory devices [6], etc. In addition, ternary compounds have wide bandgap tunability depending on the element composition [7]. Among ternary semiconductor NWs, aluminum gallium nitride ($Al_xGa_{1-x}N$) nanowires are considered very promising for ultraviolet light-emitting diodes and lasers [8–11], due to their tunable direct bandgap (3.4 eV - 6 eV, corresponding to ~207 nm - 364 nm). Moreover, AlGaN thin films and NWs have been considered for efficient electron field emission cathodes with low threshold electric field and large emission current density [12–14]. Field emission (FE) devices are very attractive to realize controlled electron transport in vacuum [15], flat panels [16,17], high frequency electromagnetic waves amplifiers and generators [18–20], electron [21] or X-ray sources [22,23]. Electron field emission from nanostructured surfaces of GaN and AlGaN has been reported [24] at turn-on fields of the order of 10-100 V/µm and explained in terms of effective work function lowering caused by electron inter-valley transition upon heating and on energy band reconstruction of the surface of the nanowires due to quantum size-confinement effect [24]. Low turn-on voltage of 2.3 V has been reported [13] for AlGaN/GaN lateral field emission nano-void channel device in which the semiconducting emitter and the metal collector are separated by a gap of about 45 nm, corresponding to a turn-on field of about 50 V/µm. Nanostructured AlGaN/GaN quantum wells have been suggested [25] as field-enhanced multilayered quantum emitters, more robust than 1D nanostructures under high current density conditions, and heve been theoretically modelled considering a generalized FE mechanism in which three distinct FE modes depending on the applied field are proposed [12]. Ultrathin layers of GaN(4 nm) / $Al_{0.5}Ga_{0.5}N$ (2 nm) have been used as electron emitting surface [26] reporting stable FE current densities up to $3\times10^{-2}$ A/cm$^2$ with a turn-on field of about 50 V/µm. In this case, a dual-barrier model with a two-step mechanism for the electron emission process is proposed [26], taking into account the space charge formation in the quantum well structure at the surface. Reproducible peaks in the field emission characteristics have been reported [27] for photoelectrochemically etched $Al_{0.3}Ga_{0.7}N$/GaN heterostructures and interpreted as resonant tunnelling through a multibarrier structure.

In this paper, we report a comprehensive study of the field emission properties of AlGaN nanowires, performed inside a scanning electron microscope vacuum chamber in which a nanometric tip-shaped anode enables the collection of electrons from areas as small as 1µm$^2$. We show the homogeneity of the field emission properties on large areas and we report evidence of low turn-on field (17 V/µm) and high field enhancement factor (556) at very small cathode-anode separation distance (500 nm). We finally demonstrate the robustness and the stability of AlGaN NWs as field emitters and their suitability for large area cold cathodes.

## 2. Materials and methods

Radio-frequency (RF) plasma-assisted molecular beam epitaxy (MBE) technique was used to grow AlGaN nanowires (NWs) on a GaN nanowire template on Si(111) substrates. Silicon solvent cleaning and hydrofluoric acid etching were employed to clean the substrate and remove the surface oxide before loading in the MBE chamber. Then, thermal outgassing was performed at 830°C before the growth process. We notice that no patterning procedure was involved in the fabrication process. First, GaN nanowire template was grown for 30 minutes at $T_{GaN}$ = 760 °C followed by 1 hour AlGaN growth at 785 °C obtaining a nanowire density of ~1.5 × 10$^{10}$ cm$^{-2}$, overall NW length of about 300 nm and diameter of about 30-50 nm. NW density can be tuned by varying the growth temperature of the GaN template, from ~0.8 × 10$^{10}$ cm$^{-2}$ (for $T_{GaN}$ = 775 °C) to ~2.4 × 10$^{10}$ cm$^{-2}$ (for $T_{GaN}$ = 745 °C). Further fabrication details are available elsewhere [28,29]. In this study, samples with 40% Al content are investigated. In figure 1a, the top image of AlGaN NWs by scanning electron microscope (SEM) is reported.

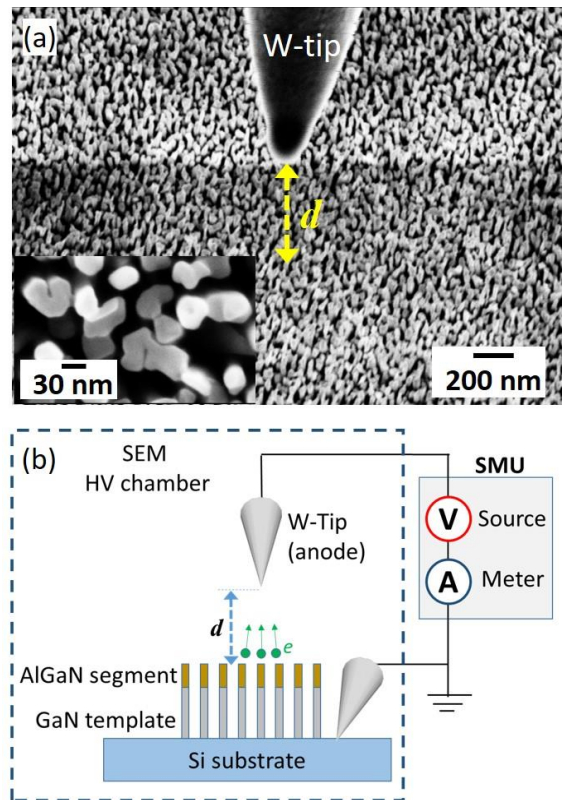

**Figure 1.** (*a*) *SEM image of AlGaN nanowires and tungsten tip anode approaching the surface at a distance d from the samle surface. Inset: magnified image of the surface. (b) Schematic of the FE setup inside the vacuum chamber of a SEM. Nano-controlled electrodes (tungsten tips) are connected to external source-meter unit.*

The crystalline structure of the sample was investigated by means of X-ray diffraction (XRD) technique, using a Bruker D5000 setup equipped with Cu K$_\alpha$ (wavelength $\lambda$ = 0.154 nm) line excitation source and in Bragg-Brentano geometry, where almost all the information about the crystal structure comes from the lattice planes parallel to the substrate surface. All the XRD spectra were acquired using a Ni filter to reduce the intensity of the $K_\beta$ line in the X-ray spectrum.

X-ray photoelectron spectroscopy (XPS) has been used to analyze the surface elemental composition of the samples under investigation. Experiments have been performed by means of PHI 1257 system equipped with a hemispherical analyser (Physical Electronics Inc.), operating at pressure lower than $10^{-9}$ mbar and using a non-monochromatic Mg $K_\alpha$ source (hυ = 1253.6 eV).

The experimental setup for field emission characterization is realized inside a Zeiss LEO 1530 scanning electron microscope (SEM) working at a base pressure of about $10^{-6}$ mbar and equipped with two piezoelectric driven nanomanipulators (MM3A by Kleindiek) to finely control (with step resolution of 5 nm) tungsten tip probes with curvature radius of about 100 nm. The metallic tips (working as in-situ electrodes) are connected to a semiconductor parameter analyzer (Keithley 4200 SCS) to perform electrical measurements by applying a bias in the range 0 – 120V and measuring the current with a resolution better than $10^{-13}$ A. The sample under investigation and the nanomanipulators are both installed on a tilting stage inside the vacuum chamber of the microscope, allowing precise estimation of the cathode-anode separation distance for the field emission device. A layout of the experimental setup is reported in figure 1b.

## 3. Results and discussion

### 3.1 XRD analysis

In figure 2a, we show the typical 2θ XRD pattern of AlGaN/GaN NWs on silicon (111) substrate, comparing it with the pattern obtained for the bare silicon (111) substrate. Si (111) and its 2$^{nd}$ order peaks (not shown here) are resolved at 28.4° and 58.9°, respectively. Moreover, the diffraction peaks for GaN(0002), $Al_{0.4}Ga_{0.6}N$(0002) and Si(111), confirm the high crystallinity with the c-axis normal to Si(111). It has been shown [30] that AlGaN layer on AlN template exhibits an XRD pattern with fixed AlN(0002) peak due to the template, while the AlGaN peak appear at angle higher than what is expected for GaN(0002) with peak position moving towards GaN(0002) position for reducing Al content. Similarly, for AlGaN NWs grown on GaN template on Si(111) substrate, we found the GaN(0002) peak at 34.66° and $Al_{0.4}Ga_{0.6}N$(0002) peak at 35.60° (between 34.63° for GaN(0002) and 36.04° AlN(0002) at which would be expected as reported in the JCPDS 065-3410 and JCPDS 025-1133, respectively). At higher angles, we also resolve GaN(0004) and $Al_{0.4}Ga_{0.6}N$(0004) peaks at 73.20° (in accordance with the JCPDS 065-3410) and 75.62°, respectively, while AlN(0004) would be expected at 76.44° (according to the JCPDS 025-1133). These data confirm the high homogeneity of Al doping in the NWs, with no indications of Al segregation or pure AlN formation, and that there is no presence of cubic phase of GaN, within the detection limit of the XRD, thus it can be confirmed that the $Al_{0.4}Ga_{0.6}N$ and GaN NWs have a hexagonal structure. As further confirmation, we show in figure 2a the comparison with the XRD pattern taken on a sample with a lower Al content (~20%), grown with the same process. Accordingly, the GaN related peaks appear at the same angles as for 40% Al content, while the $Al_{0.2}Ga_{0.8}N$ related peaks, due to the reduced doping, appear shifted to lower angles, closer to the GaN peaks. The peak position measured from the XRD spectra allows us to calculate the lattice constant $c$ by using Bragg's law: $d(hkl) = \lambda/2sin\theta$ and $d(hkl) = c/\sqrt{h^2 + k^2 + l^2}$, where $\lambda$ is the wavelength of X-ray and $d(hkl)$ is the distance between planes given by the Miller indices ($hkl$). The lattice constant $c$ was calculated from all the symmetric reflections and the corresponding values have been listed in Table 1.

**Table 1.** *Summary of the lattice constant c obtained by XRD spectra.*

| NWs | $2\theta°$ | $d(0002)$ (Å) | Lattice constant $c$ (Å) |
|---|---|---|---|
| GaN | 34.66 | 2.586 | 5.172 |
| $Al_{0.4}Ga_{0.6}N$ | 35.60 | 2.520 | 5.040 |
| GaN | 34.63 | 2.588 | 5.176 |
| $Al_{0.2}Ga_{0.8}N$ | 35.02 | 2.560 | 5.120 |

### 3.2 XPS characterization

XPS measurements are very useful to determine the chemical composition of the surface, giving information on the elements as well as their chemical state, for a thickness of about 10 nm. In figure 2b we show the survey spectrum, measured for binding energies (BE) up to 1200 eV, in which we can observe several peaks related to Ga, Al, C, N, and O. We have then calibrated all XPS spectra with respect the C 1s peak, considering that the binding energy of adventitious carbon is expected 284.8 eV [31]. Moreover, all peaks were deconvoluted, using Voigt multipeaks and a Shirley background. In Figure 2c, we show the spectrum N 1s that is the result of convolution of two main component peaks centered at BE of 397.0 eV (N-Al bond) and 398.3 eV (N-Ga bond), showing a quite large area ratio of N-Al to N-Ga, in agreement with previously reported data for AlGaN with similar or even lower Al content [32,33], probably due to a slightly excess of Al content in the top most surface layer. In Figure 2d, we show the Al 2p spectrum that can be reproduced by two component peaks corresponding to Al-N bonds at a BE= 73.4 eV and Al-O bonds at 74.5 eV, the area ratio Al-O/Al-N being about 0.17.

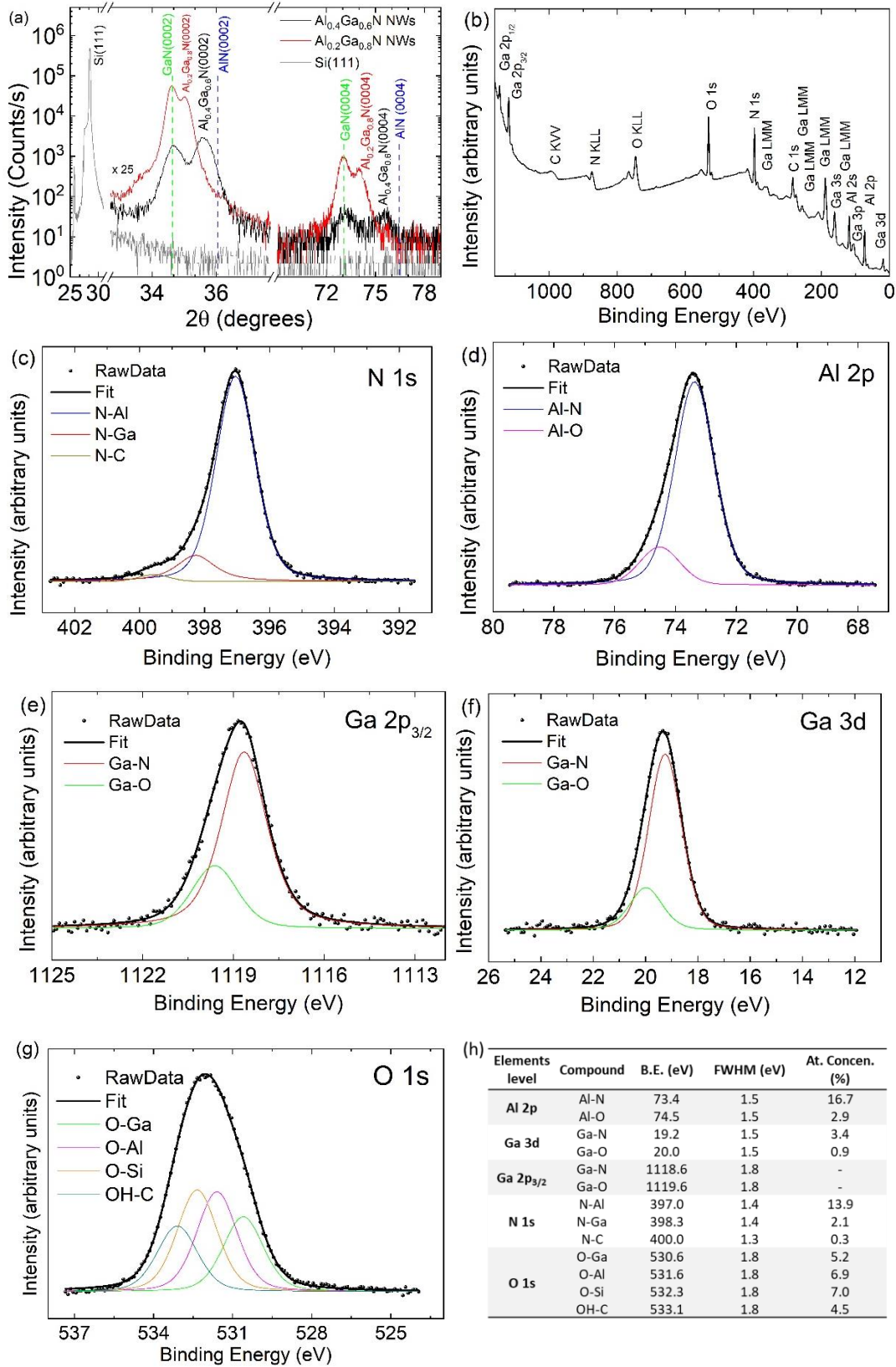

**Figure 2.** *XRD and XPS characterization of AlGaN nanowires. (a) XRD pattern, measured for $Al_{0.4}Ga_{0.6}N$ nanowires grown on GaN template on Si (111), is compared to XRD pattern of the $Al_{0.2}Ga_{0.8}N$ nanowires and of the Si (111) substrate. (b) XPS characterization, survey spectrum; deconvoluted XPS spectra of (c) N 1s, (d) Al 2p, (e) Ga $2p_{3/2}$, (f) Ga 3d, (g) O 1s. (h) Summary of XPS data: BE, FWHM and atomic concentration.*

Ga 2p$_{3/2}$ spectrum is deconvoluted in Figure 2e, evidencing the presence of two component peaks, namely the Ga-N bonds at 1118.6 eV and the Ga-O bonds at 1119.6 eV [33]. Finally, Ga 3d spectrum is deconvoluted (Figure 2f) in two component peaks as well, the Ga-N bonds at 19.2 eV and the Ga-O bonds at 20.0 eV, with an area ratio Ga-O/Ga-N of about 0.26. The signal due to Ga-O bonds (and to Al-O bonds) indicates the presence of a thin oxide layer on the surface, which is probably amorphous, not being evidenced by the XRD pattern. We notice that a thin oxide layer (of the order of 1 nm) does not significantly affect the FE behavior of the NWs, the only effect being to introduce a small extra tunnel barrier to overcome, corresponding to a slightly increase of the turn-on field with respect of an oxide-free surface.

In figure 2g, we also show the deconvolution of the O 1s spectrum, identifying four component peaks, namely the O-Ga bonds at 530.6 eV, the O-Al bonds at 531.6 eV, the O-Si bonds at 532.3 eV and the OH-C bonds at 533.1 eV.

In figure 2h we summarize the BE, the Full Width at Half Maximum (FWHM) and the atomic concentration for the deconvoluted XPS spectra.

### 3.3 Field emission

To perform field emission characterization, we started by approaching both nano-manipulated probes on the sample surface. Once the electrical contact was established, the tungsten tip used as the anode was gently detached and positioned at controlled distance from the surface, thus obtaining the circuit configuration for FE measurements (schematic is shown in figure 1b). The piezoelectric control of the tungsten tips allows precise tuning of the cathode-anode separation distance $d$ with 5 nm spatial resolution. In figure 3a we show the current-voltage ($I-V$) characteristics, measured by applying an external bias sweep up to +100V on the tip anode, to collect electrons field emitted from the AlGaN nanowires, for different cathode-anode separation distances, in the range 500 nm – 1400 nm. Larger bias was avoided to prevent discharge failure, emitters burning and/or damages to the nano-manipulator circuitry. In order to verify the robustness of the FE device as well as to check the repeatability of the measurements, we standardly performed at least two successive I-V sweeps; this allowed us to exclude effects and/or modifications induced by electrical stress. Similarly, we repeatedly performed SEM imaging of the emitting area under investigation, before and after I-V measurements, to further confirm the surface stability.

In figure 3b, experimental data are plotted on logarithmic scale in order to better visualize the current rising. Indeed, for $d = 500$ nm, we observe a turn-on voltage (here defined as the voltage necessary to get a current of 1 pA) $V_{on} = 15$V. Then, the current exponentially grows for about six orders of magnitude up to the µA. Similarly, at $d=$ 1100nm the FE current starts at $V_{on} = 38$V and at $d=$ 1400nm the FE current starts at $V_{on} = 52$V. To analyse the I-V curves, we compare experimental data with the expected behaviour (solid lines in figure 3b) in the framework of Fowler-Nordheim (FN) theory [34]. According to this model, the emitted current from a planar metallic surface can be expressed in the form:

$$I = S \cdot A\phi^{-1}\left(\beta\frac{V}{d}\right)^2 exp\left[-B\ \phi^{3/2}\left(\beta\frac{V}{d}\right)^{-1}\right] \quad (1)$$

where $S$ is the effective area from which electrons are field emitted, $A= 1.54 \times 10^{-6} AV^{-2}eV$ and $B= 6.83 \times 10^9 eV^{-3/2}m^{-1}V$ are dimensional constants, $\phi = 4.93$ ev [35] is the work function; $V/d = E$ is the electric field due to the applied voltage $V$ when the anode-cathode separation distance is $d$, while β represents the field enhancement factor that originates the amplification of local electric field $E_L$ in proximity of curved surface or apex ($E_L = \beta V/d$). Accordingly, the ln(I/V$^2$) versus 1/V (the so-called FN plot) is expected to be a linear function, and from the slope $m$, one can calculate the field enhancement factor as $\beta = -Bd\phi^{3/2}m^{-1}$. We remind here that equation (1) has been derived for standard parallel plate configuration [36] in which a planar anode is faced to a metallic and planar emitting surface.

However, despite such FN theory has been developed for emission from flat metallic surface under a uniform macroscopic electric field, it is widely accepted that it can be used as first approximation to identify the field emission nature of current collected from nanostructured surfaces, in presence of non-uniform electric field. Indeed, FN theory has been successfully applied to analyse FE from nanoparticles [37–40], semiconducting nanostructures [41], nanowires [42–46], nanotubes [47–54], two-dimensional nanosheets, such as graphene [55–58] and transition metal dichalcogenides [59–68]. For the sake of completeness, we mention here that corrections have been discussed in the literature [69–74] to include effects due to curved surfaces, series resistance, non-homogeneous field enhancement factor or work-functions, one-dimensional and two-dimensional nanostructures, semiconducting emitters, electronic structure of emitters, etc. [70–72,75]

In the following, for a more precise data analysis we need to take into account that in our setup we are employing a tip-shaped anode with a small curvature radius of about 100 nm. It has already been demonstrated that in such configuration, the field emission current collected at the anode comes from a limited circular area having diameter of the order of the cathode-anode separation distance. [76] Moreover, the anode geometry modifies the distribution of the electric field lines and the effects on the I – V characteristics can be easily modelled by introducing a correction factor $k_{tip} \approx 1.6$ that takes into account the anode geometry, [76] so that $\beta =$ $-Bk_{tip}d\phi^{3/2}m^{-1}$. Moreover, the use of a tip-shaped anode allows to collect field emitted electrons from limited sample areas as small as ~1μm² and thus to obtain local information on the FE properties [76], in comparison to the standard parallel plate setup that provides information on areas up to mm².

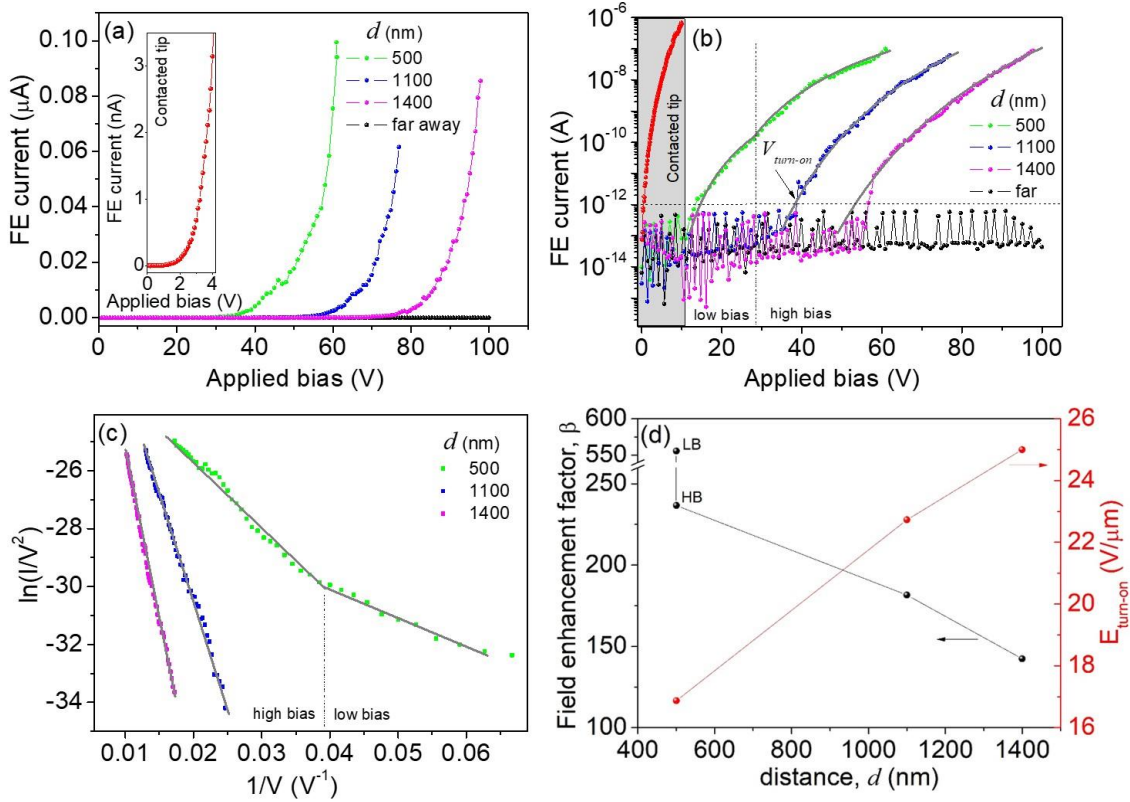

**Figure 3.** *(a) Field emission I-V characteristics for various cathode-anode separation distances plotted on linear scale. Inset shows the curve when anode tip is in contact with the sample surface. (b) I-V characteristics plotted on semi-logarithmic scale are compared to numerically calculated curves obtained according to Fowler-Nordheim model (Equation 1). Horizontal dotted line evidences 1pA level. (c) FN-plots corresponding to the I-V curves and linear fittings (solid lines). (d) Field enhancement factor β and turn-on field $E_{turn-on}$ as a function of the cathode-anode separation distance d. For d = 500 nm, both β values, for low bias (LB) and high bias (HB), are reported.*

In figure 3c, we show the FN plots corresponding to the experimental data of figure 3b. Solid lines represent the linear fittings. The linearity of the FN plot demonstrates that the measured I-V curves are governed by the conventional FN tunneling. Thus, we can estimate the field enhancement factor for each value of the cathode-anode separation distance, by using the slope of the FN plot obtained from the numerical fitting. We notice that the field enhancement factor results to be a decreasing function of $d$ in the investigated range 500 nm $< d <$ 1400 nm (see figure 3d), the lowest value being $\beta = 142$ for $d = 1400$ nm. For the minimal separation distance, $d = 500$ nm, we observe that the FN-plot is characterized by two different slopes, $m_H = -252$ corresponding to a field enhancement factor $\beta = 237$ for the high bias region (V > 30 V) and $m_L = -107$ corresponding to a field enhancement factor $\beta = 556$ for the low bias region (V < 30 V). Similarly, we plot in figure 2d the dependence of turn-on field $E_{turn-on}$ (here defined as the electric field necessary to obtain a FE current of 1pA) on the separation distance $d$ and we observe that it is an increasing function, from a minimal value $E_{turn-on} \approx 17$ V/μm at $d = 500$ nm to a maximum value $E_{turn-on} \approx 25$ V/μm at $d = 1400$ nm. We notice here that the tip-shaped anode also affect the calculation of the turn-on field, indeed $E_{turn-on} = k_{tip}^{-1}V_{on}/d$ [76].

The FE characterization has been repeatedly performed in several different locations of the sample surface, in order to verify the homogeneity of FE properties, and the suitability of the samples for large area applications. We show, as comparison, in figure 4, the experimental data obtained in another location. The I-V curves have been measured for several separation distances in the range 500 nm $< d <$ 1600 nm (figure 4a). Once again, the FE current is a fast increasing function of the externally applied voltage and it is well described by the FN theory, as confirmed by the linear FN plots reported in figure 4b. From the slope of the linear fittings, we can calculate the field enhancement factor for the various separation distances in this new location.

We have found that for the minimum separation $d = 500$ nm, a double slope in the FN-plot is observed and we estimate a field enhancement factor $\beta = 570$ for the low bias region and $\beta = 225$ for the high bias region, with $E_{turn-on} \approx 18$ V/μm. Similarly, for the maximum separation $d = 1600$ nm, we obtain $\beta = 105$ and $E_{turn-on} \approx 34$ V/μm. The data are then compared to the values reported for the first location in figure 4c. Interestingly, we observe very similar behaviour for β vs $d$ (decreasing) and $E_{turn-on}$ vs $d$ (increasing).

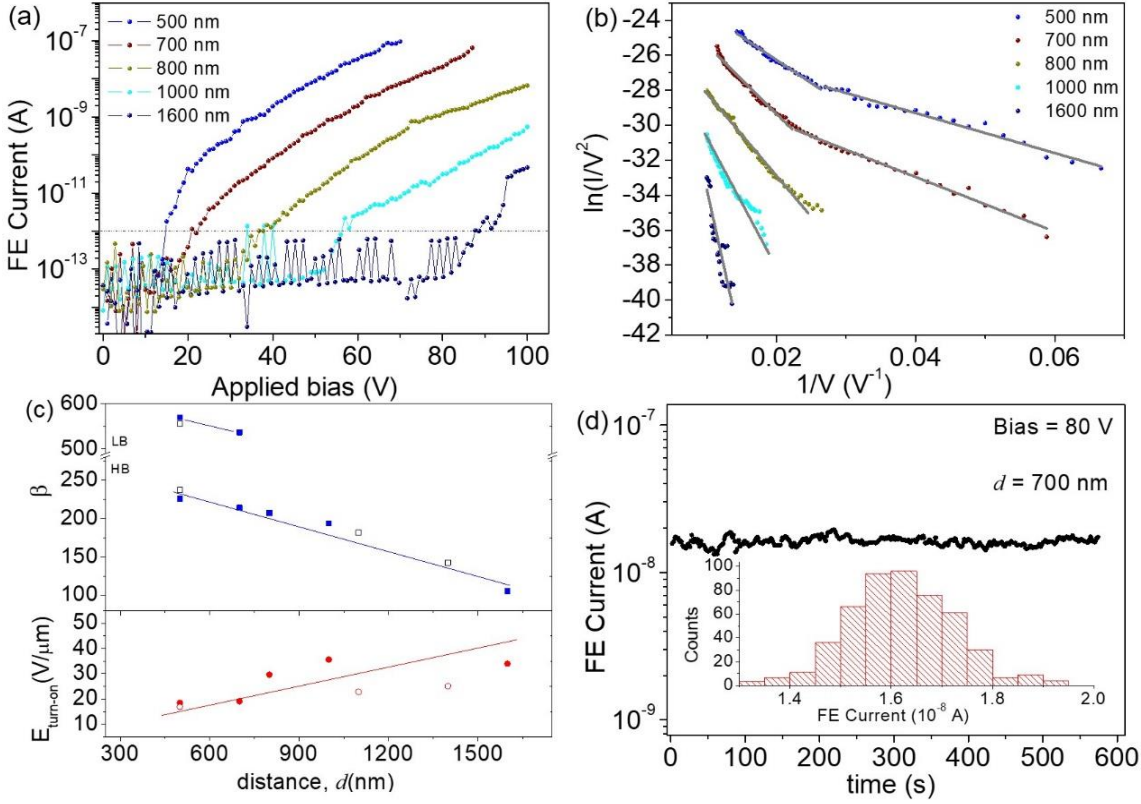

**Figure 4.** *(a) Field emission I-V characteristics measured in a different location of the sample surface. Horizontal dotted line evidences 1pA level. (b) FN-plots corresponding to the I-V curves. (c) Field enhancement factor β (upper plot) and turn-on field $E_{turn-on}$ (lower plot) as a function of the cathode-anode separation distance $d$ (full symbols) are compared to previously reported values (empty symbols). Solid lines are guides for eyes. (d) Current stability (FE current versus time) measured at a cathode-anode separation distance d=700nm and at a constant bias of 80 V. The inset represents the histogram obtained from the statistical analysis of the recorded current values.*

The time stability of the FE current is another crucial characteristic of a good emitter, together with the turn-on field and the field enhancement factor. Consequently, we measured the FE current versus time by keeping a fixed bias on the tip anode. We performed the experiment positioning the anode at $d = 700$ nm from the sample surface and by applying a bias of 80 V. Experimental results are reported in figure 4d. We observe a stable current over several minutes without any significant current degradation. Moreover, from a statistical analysis of the current values (see inset of figure 4d) we notice that current fluctuations (around the average value of 1.6·10⁻⁸ A) are as small as 1nA (standard deviation).

We notice that in both locations, I-V characteristics show a change of slope in the FN plot, when FE is starting already at low bias. More precisely, in figure 3b we observed that the curve measured at $d=$ 500 nm has a low turn-on voltage of 17 V and the numerical simulation (according to FN theory) needs to use different parameters for low bias (LB<30V) and for high bias (HB>30V). It is not possible to reproduce experimental data with unique curve modelled by equation (1). However, in both regions the I-V curves follow separately the expected behaviour in the FN model. This is observed also in the FN plot of figure 3c, where we clearly see the change of slope in the linear behaviour. Similarly, the change of slope in the FN plots have been found for curves measured in different sample locations, as reported for example in figure 4b. To explain the existence of two different emission regimes, one at low bias corresponding to a larger field enhancement factor, and one at high bias corresponding to a reduced field enhancement factor, we need to take into account that we are in presence of a large number of emitters that are not identical. In particular, the different heights (as observed by SEM imaging) will originate emitters with different enhancement factor. Consequently, at low applied voltages, only protruding emitters (i.e. with larger β) will contribute to the emitted current, originating the first part of the I-V characteristic with a fast raising branch. For larger applied voltages, also emitters with reduced β will start the emission process. Consequently, at high bias, the contribution to the FE current will be due to both kind of emitters, and the average behaviour will correspond to a lower field enhancement and thus to an increased slope of the FN plot as reported in our experimental data. A similar behaviour, with deviation of FN plots from the linearity, has been reported for graphitic nanocones [77]. Also in that case, experimental results and numerical calculations demonstrated that such behaviour was originated from the non-uniform field enhancement factors of the emitters. At low field, electrons were emitted mainly from nanocones with large β (corresponding to the FN plot with small slope), while at strong field, nanocones with small β also started to emit, reducing the average β (corresponding to an increased slope of the FN plot).

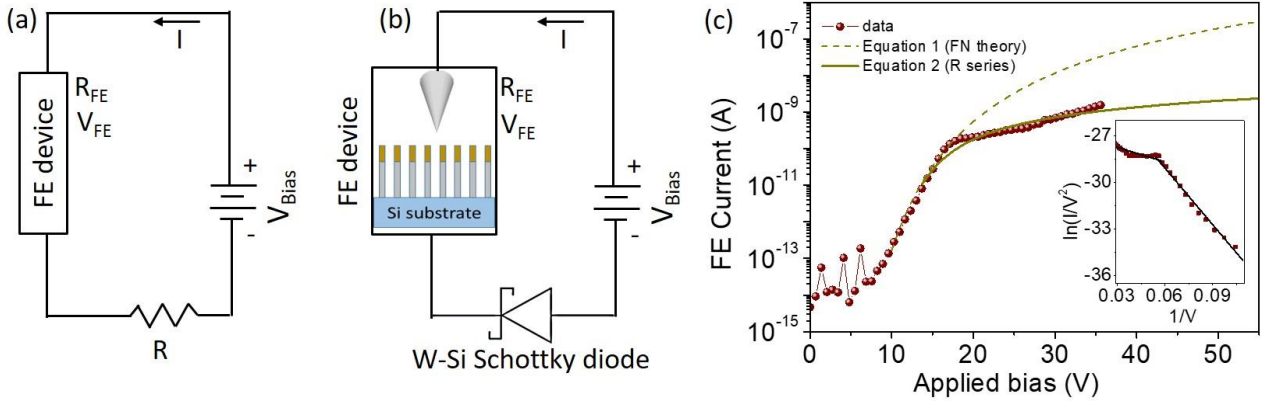

**Figure 5.** *(a) Schematic of the FE circuit including a series resistance R. (b) Schematic of our FE setup with Schottky diode formed at the W-Si contact. (c) FE I – V characteristic measured after realizing a bed contact between the tip cathode and the sample in order to produce high contact resistance in series with the FE device. Experimental data are compared to FN theory (equation 1) and to the model (equation 2) that includes the presence of the series resistance R. Inset: FN plot.*

In some cases, it has been found that the change of slope in the FN plots could originate instead from the presence of high resistance in series in the field emission circuit [76,78]. Indeed, from the schematic of the FE measurement setup reported in figure 5a, it is easy to understand that the presence of a large resistance R in series with the FE device causes a reduction of the applied voltage $V_{Bias}$ (due to the voltage drop on R). As a consequence, the voltage $V_{FE}$ applied on the FE device is $V_{FE} = V_{Bias} - RI$. Assuming that the resistance R is not dependent on the field, and considering the expression of the current I according to equation 1, we obtain a recursive equation that we can solve numerically, in which the applied voltage is replaced by the effective voltage, $V - RI$:

$$I = S \cdot A\phi^{-1} \left(\beta \frac{V-RI}{d}\right)^2 exp\left[-B\, \phi^{3/2} \left(\beta \frac{V-RI}{d}\right)^{-1}\right] \quad (2)$$

In our setup, a second tungsten tip is used as cathode to contact the silicon substrate. We typically obtained ohmic contact that did not affect the field emission device, as shown in our experimental data reported above. On the other hand, in few cases, the tip pressure on the semiconducting surface caused the formation of a Schottky contact (see figure 5b). Indeed, a metallic (M) tip pressed on a semiconducting (S) surface forms a MS point contact, in which a Schottky barrier can be formed at the interface. In this case, the MS contact behaves as a (Schottky) diode, and no charge carriers flux is expected through the barrier when the device is reverse biased [79,80]. During FE measurements, anode tip is positively biased, and consequently the W-Si Schottky diode results reversely biased. This causes an increase of the depletion region that would stop the current flow, unless a small leakage current. Increasing the reverse bias voltage, the current also gradually increases due to the weak barrier.

In figure 5c, we report an FE characteristic, I – V, that we measured when the W-Si contact inadvertently formed a Schottky diode. We can clearly observe that after the turn-on voltage, the FE current starts to raise according to FN theory (equation 1). However, at certain voltage, experimental data sensibly deviate from the expected FN behaviour (dashed line in figure 5c), showing limited current for voltages greater than 20 V. As already stated above, we notice that the positive biasing of the anode tip corresponds to reverse biasing of the Schottky diode. In turn, the behavior of such a diode in reverse bias can behave like a very high resistance. Consequently, we can use equation 2 to simulate experimental data reported in figure 5c, assuming a series resistance of few GΩ in series in the FE circuit.

Finally, we verified that FE curves reported for AlGaN NWs and showing two slopes in the FN plots (experimental data reported in figure 3 and 4) cannot be reproduced by equation 2, in that case having an ohmic contact at the W-Si junction.

## 4. Conclusions

The field emission properties of AlGaN nanowires, grown by molecular beam epitaxy on GaN template on Si (111) substrate, have been investigated in detail. Piezo-driven tip-shaped anode (with curvature radius ~100 nm) is employed to finely control its position at separation distances from the emitters of hundreds nanometres, thus allowing to collect emitted electrons from area as small as 1µm$^2$. Current-voltage characteristics are analysed in the framework of Fowler-Nordheim theory, evidencing a low turn-on field of 17 V/µm and a field enhancement factor as high as 556, at a cathode-anode separation distance of 500 nm. We explain the observation of two different slope in the F-N plots, as the existence of two different regimes, one at low fields in which only protruding emitters contribute to the FE current, a second one at higher fields in which also shorter emitters start to emit electrons, causing an average lower value of the enhancement factor.